# Superradiant diamond color center arrays coupled to concave plasmonic nanoresonators

Dávid Vass[1,*], András Szenes[1,*], Balázs Bánhelyi[2], Tibor Csendes[2], Gábor Szabó[1], Mária Csete[1,✉]

[1]Department of Optics and Quantum Electronics, University of Szeged, Dóm tér 9, Szeged 6720, Hungary
[2]Department of Computational Optimization, University of Szeged, Árpád tér 2, Szeged 6720, Hungary
*authors equally contributed
✉ mcsete@physx.u-szeged.hu

Different types of concave plasmonic nanoresonators have been optimized to achieve superradiantly enhanced emission of SiV color centers in diamond. Comparative study has been performed to consider advantages of different $N$ number of SiV color centers, different diamond-silver (bare) and diamond-silver-diamond (coated) core-shell nanoresonator types, as well as of spherical and ellipsoidal geometry. The complete fluorescence enhancement (qualified by $P_x$ factor) monitoring and the $cQE$ corrected quantum efficiency weighted $P_x cQE$ objective function optimization promotes to design bad-cavities for plasmonic Dicke effect. The switching into a collective Dicke state via optimized nanoresonators results in a radiated power proportional to $N^2$, which manifest itself in an enhancement proportional to $N$ both of the excitation and emission rates. Accordingly, enhancement proportional to $N^2$ of the $P_x$ factor and $P_x cQE$ has been reached both via four and six SiV color centers arranged in symmetrical square and hexagonal patterns inside all types of inspected nanoresonators. Coated spherical and bare ellipsoidal nanoresonators result in stronger non-cooperative fluorescence enhancement, while superradiance is better achieved via bare spherical nanoresonators independently of SiV color centers number, and via coated (bare) ellipsoidal nanoresonators seeded by four (six) SiV color centers. Indistinguishable superradiant state of four color centers and line-width narrowing is achieved via bare nanoresonators. Six color centers seeded bare spherical (ellipsoidal) nanoresonators result in larger fluorescence enhancement and more significantly overridden superradiance thresholds, while having slightly more (less) pronounced bad-cavity characteristics. Both phenomena are simultaneously optimized in ellipsoidal bare nanoresonators embedding six color centers with a slightly larger detuning.

## 1   Introduction

The superradiance (SR) predicted first by Dicke has been thoroughly studied throughout the last half-century [1, 2]. In case of cooperativity $N$ emitters can exhibit $N$-times shorter radiative decay, accordingly the maximum rate of emission can be proportional to $N^2$. Contradicting predictions also appeared in the primary literature regarding that the coherence can be lost within the expected superradiance lifetime caused by spatially varying frequency shift [3]. However, the principles regarding that Dicke effect can be achieved either via systems, which are initially in excited "collective Dicke" state, or via transient amplification of the photon noise, are widely accepted by the scientific community [4, 5]. In sub-wavelength emitter arrays the non-uniform distribution of initial phases is the pre-condition of a superradiant burst [6]. The SR phenomenon has been widely inspected in case of various systems, which are either significantly smaller or larger than the wavelength [7]. Among large-scale superradiant systems the slab geometry with half-wavelength-scaled thickness was thoroughly studied [8]. A particularly important fundamental phenomenon is the Dicke quantum phase transition inside an optical cavity [9, 10, 11].

An emerging research area is the investigation of the plasmonic Dicke effect (PDE). The simplest system that exhibits PDE is an array of emitters around a solid plasmonic particle [12, 13, 14]. In primary studies the phase of the dipolar emitters was supposed to be random, while their orientation was uniformly perpendicular to the solid plasmonic nanoparticle interface, and both random spatial distribution [12] and a fullerene like spherical lattice [13, 14] were considered.

Both the direct coupling through radiation and the indirect coupling through plasmons were taken into account. The main conclusions were that the plasmon assisted coupling overrides the direct radiative coupling between emitters, the total radiative rate is proportional to $N/3$, while the total energy is thrice of the individual emitter energy. The plasmon mediated cooperative coupling phenomena were analyzed in similar systems consisting of a spherical metal core covered by a gain medium shell, by taking into account all mutual couplings and by treating the dye via an equivalent polarizability [15].

The effect of a metallic nanoparticle embedded into symmetric arrays of already collectively oscillating dipolar emitters was inspected as well [16]. However, to achieve a collective Dicke state the indistinguishability of the emitters manifesting itself in synchronized phases was required. In these studies, the dipolar emitters were aligned parallel and equidistantly, which ensured that only symmetric states could be at play and only one single plasmon could be excited. It was demonstrated that the nanoparticle accelerates the superradiance, decreases the SR pulse delay as well as its duration. The Purcell-Dicke effect of an atomic ensemble near a conducting surface was qualified by the Purcell fidelity that exhibits a superradiant burst in its dynamics [17]. It was demonstrated that the metal surface mediated collective interactions can lead to superradiance (subradiance) in case of constructive (destructive) interference between emitters [18]. Switching into plasmon-mediated superradiant state via cooling was attributed to polarization phase matching of the molecular transitions in emitters embedded into a dielectric shell around a gold core [19]. It is important to notice that in all previous examples of plasmonic superradiance -except the first three cases- all emitters are in an excited, moreover in synchronized collective Dicke state [15-19].

The Purcell-Dicke effect was demonstrated in concave spherical nanoresonators (NRs) as well, namely in an ensemble of resonant atoms, which are embedded into a spherical dielectric core coated by a metallic shell [20, 21]. Existence of one (two) superradiant modes was proven by neglecting [20] (taking into account [21]) the wavelength dependency of the optical parameters. Moreover, an array of cylindrical concave NRs embedded into a metal film was also applied to generate SR. It was shown that a continuous wave plasmonic superradiance is achievable from a 2D spaser array due to the synchronization of plasmonic nanoholes via gain molecules [22]. Narrow beam of intense coherent light can be extracted into the far-field from such an array.

There are several important applications of the photonic superradiance, among them is design of superradiant lasers, which exhibit ultra-narrow lines [23]. It has been demonstrated that the quantum measurement precision can be improved via squeezed Dicke states [24]. Dicke superradiance in small e.g. two qubits ensembles makes it possible to explore different aspects of entanglement [25]. There are emerging applications of the plasmonic superradiance as well. An ordered radial arrangement of few dipolar emitters around a solid plasmonic sphere was inspected as a potential candidate for multi-qubits deterministic quantum phase gate [26]. In this approach the inter-emitter interactions have been ignored, while the plasmon mediated long-range interactions have been taken into account. In Ag doped oxyfluoride, where the wide nanoparticle and the size-dependent cluster band's overlap mediates the Ag nanoclusters interaction, the plasmonic Dicke effect can be used to generate picosecond pulses [27].

Diamond is a promising candidate medium, since superradiance was already demonstrated from NV centers in diamond nanocrystals [28]. Namely, it was shown that the bright diamond nanocrystals are faster, and decay rates in the order of ~1 ns were reported. Although, the value of the second order correlation function was initial spin population dependent, it proved the cooperative nature of the diamond nanocrystal emission in certain configurations. Moreover, cooperatively enhanced trapping was also demonstrated in case of nanodiamonds consisting of large density NV centers [29]. SiV is a particularly promising diamond color center due to the achievable larger density, corresponding stronger transition moment and the resulted narrow fluorescence line [30, 31, 32]. Both in bulk and in nanofabricated diamond consisting of ion implanted SiV, the almost lifetime limited line-width is accompanied by an extremely small inhomogeneous distribution, which is the precondition to ensure indistinguishable photons [32].

## 2  Methods

Concave spherical and ellipsoidal core-shell nanoresonators have been optimized to maximize the fluorescence rate from various arrays of multiple SiV color centers treated as dipolar emitters by using a FEM (COMSOL Multiphysics) based method described in our previous publications (Fig. 1-5, SFig. 1-5, STable 1-4) [33, 34, 36]. Considering that symmetry breaking causes different close-neighboring environment, dephasing and frequency chirp, while indistinguishability of emitters is an important peculiarity of Dicke effect, $N = 4$ and 6 SiV color centers arranged in symmetric arrrays have been inspected and compared [4].

Accordingly, the inspected spherical (ellipsoidal) NRs consist of a symmetrical square or hexagonal pattern of 4 or 6 SiV color centers, which oscillate in the equatorial plane (corresponding to the short axis) at the excitation and perpendicular to it (along the long axis) at the emission, according to SiV color center transition dipole's perpendicularity. In case of synchronization 4 emitters in excitation configurations and both of 4 and 6 emitters in emission configurations are indistinguishable in all NRs due to the symmetry properties of their arrays. In contrast, two subsystems of 2 and 4 emitters located at 0° and 60° azimuthal orientation with respect to the x axis in excitation configuration are distinguishable in NRs seeded by 6 SiV color centers. The diamond-silver core-shells standing in air are referred to as bare type spherical and ellipsoidal NRs, while the core-shells of same composition covered by an additional diamond layer are referred to as coated type spherical and ellipsoidal NRs (SFig. 1a, Fig. 2).

Conditional optimizations have been performed via an in-house developed algorithm integrated into COMSOL software package, by setting a criterion regarding the radiative rate enhancement at the excitation, which equals to the product of the *Purcell factor* and the quantum efficiency at the specific wavelength ($\delta R_{ex}$=*PurcellQE*) [33-36]. Accordingly, only those systems were evaluated, which ensured excitation enhancement as well. Evaluation of different systems received via conditional optimization made it possible to compare the advantages of different number of SiV color centers, as well as of different types and geometries of NRs (SFig. 1-5, STable 1-4). The optimization was performed by selecting $P_x cQE$ as the objective function, accordingly the FOM of superradiant coupled systems was the product of radiative rate enhancements at the excitation ($\delta R_{ex}$) and emission ($\delta R_{em}$), nominated as $P_x$ *factor*, and the quantum efficiency at the emission (*cQE*), which is corrected by the 10% intrinsic quantum efficiency of SiV color centers [33, 34, 36].

The optimized NRs were primarily characterized by their scattering (*scs*) and extinction (*ecs*) cross-section determined via polarized plane wave illumination (Fig. 1a, SFig. 4a-c, STable 4). Ellipsoidal NRs were illuminated by a plane wave with a polarization along their short and long axis to determine the optical cross-sections corresponding to the excitation and emission configurations, respectively. To qualify bare and coated type spherical and ellipsoidal NRs seeded by 4 and 6 SiV color centers, the wavelength dependent quantum efficiency (*QE*), *Purcell factor* and radiative rate enhancement ($\delta R$) quantities have been determined (Fig. 1b, c, SFig. 2a-c, STable 2). The $\delta R^N$ radiative rate of $N$ SiV color centers collectively oscillating inside NRs was determined by comparing their $P^{N*}$ radiated power to the $N$-fold of the $P^1$ power radiated by a single SiV color center oscillating in vacuum, i.e. existence of initially randomly oscillating color centers was supposed (SFig. 3ab and ad, STable 3a). Selection of randomly oscillating SiV color centers in vacuum as a reference system ensures that the optimized NRs result in superradiantly enhanced emission, rather than an enhanced superradiance, which has been previously described in the literature [16]. To quantify and compare different optimized coupled systems the $P_x$ *factor* and $P_x cQE$ was depicted as a function of system type (SFig. 2d). The surface charge density and near-field distributions as well as the far-field radiation pattern were inspected both at the SiV color center 532 nm excitation and 737 nm emission wavelengths (Fig. 2 and Fig. 3, SFig. 2b and c, STable 2). To prove that superradiance is achievable via $N$ color centers in the optimized configurations, the $\delta R^N$ radiative rate enhancements both at the excitation and emission were compared to corresponding $\delta R^1$ of the reference systems consisting of one single SiV color center inside a NR having the specific geometry (Fig. 4a, SFig. 3a, 3bb and cb, STable 3b&c). The optical responses of reference systems were calculated by taking the average of responses in all distinguishable, i.e. geometrically different, SiV color center positions (SFig. 3ac). Namely, the optical response in one single color center position was determined in both configurations of $N$ = 4 centers and in the emission configuration of $N$ = 6 centers in both types of spherical and ellipsoidal NRs. In contrast, the optical responses in the two distinguishable, i.e. geometrically different single color center positions were averaged in the excitation configuration of either bare or coated type spherical and ellipsoidal NRs seeded by $N$ = 6 color centers. For the sake of completeness, the ratio of the $P_x$ *factors* and the ratio of the FOMs (i.e. $P_x cQE$) was also determined, moreover the *cQE* corrected quantum efficiency achievable via $N$ color centers and via single center were compared as well (Fig. 4a, SFig. 3ba, 3ca, 3bc, 3cc, STable 3b and c). To conclude about the superradiance the ratios with respect to the reference system were normalized as follows: $r\delta R_{ex}/N$, $r\delta R_{em}/N$, $rP_x/N^2$, $rP_x cQE/N^2$ and $rcQE/1$, where r$X$ refer to the ratios of specific quantities, $N$ indicates the number of embedded color centers. That means we consider $N$ and $N^2$ as the balanced partial and complete multiplied thresholds of superradiance, respectively, however these thresholds are equivalent with the usual criterion of superradiance regarding the radiated power, which has to be proportional to $N^2$, as it is described in Equations (1) and (2) of the Supporting Information (SFig 3a, STable 3a). Finally, the average of the normalized ratios ($\overline{rX}$ =Σ($r\delta R_{ex}/N$, $r\delta R_{em}/N$, $rP_x/N^2$, $r(P_x cQE)/N^2$, $rcQE/1$)/5) was determined, to evaluate the superradiance overriding on the average.

In order to qualify the plane-wave illuminated and *N* color center seeded optimized NRs, the FWHM of the extinction and scattering cross-section peaks was determined and compared to the FWHM of the *Purcell factor* and *δR* radiative rate enhancement peaks (Fig. 4b, SFig. 4a, STable 4). The quality factor (*Q factor*) of the optimized nanoresonators was determined based on extinction cross-section spectra extracted from plane wave illumination of NRs with the specific geometry as well as based on the *Purcell factor* spectra (Fig. 4b, SFig. 4b, STable 4). To qualify the system of cooperatively oscillating emitters, detuning (*Δλ*) of the extinction and scattering cross-section peaks, *Purcell factor* and *δR* radiative rate enhancement peaks from the SiV color center emission wavelength as well as relative difference with respect to theoretical frequency pulling (*δf*) were determined (Fig. 4c, SFig. 4c and d, STable 4) [21].

The achievable $P_x$ *factor* and $P_x cQE$ quantities were compared for different number of color centers, different NR types and geometries (Fig. 4d, SFig. 2d, STable 2). To analyze the indistinguishability of *N* color centers, degeneracy in *Purcell factor* was inspected at the excitation and emission wavelengths as well (Fig. 5a, b, SFig. 5). Comparison of $P_x$ *factor* and FOM$P_x cQE$ achievable via systems consisting of collectively and randomly oscillating color centers inside the optimized NRs was also performed (Fig. 5c, d).

All comparative statements are presented in 4-to-6 SiV color centers, bare-to-coated NR type, ellipsoidal-to-spherical NR geometry sequence, and in case of emission comparative statements with respect to the excitation are also presented.

All quantities presented in the paper qualify coupled systems consisting of *N* color centers. In order to differentiate the coupled systems and the reference systems consisting of one single center, the corresponding quantities are distinguished as $X^N$ and $X^1$ in the Supporting Information (Eq. (1) and (2), SFig. 3a, STable 3a, STable 3b).

# 3  Results and discussion

The appropriately large distances between SiV color centers in the optimized NRs allow neglecting their direct coupling, similarly to previous studies in the literature [26]. The geometrical parameters (*R* core radii, *t* shell thickness, *d* SiV color center distance from shell) are very similar, when 4 and 6 SiV color centers are embedded into the same bare or coated types of optimized NRs (SFig. 1a and b, STable 1). The similar characteristics in case of 4 and 6 centers reveal that by increasing the number of embedded emitters the geometrical parameters of the optimized NRs converge.

The absence of additional shell has a well-defined effect, namely the core radius (long axis) is larger, the shell thickness is smaller, accordingly the generalized aspect ratio ($GAR=R_{1/2}/(R_{1/2}+t)$), which is the ratio of inner and outer radii) is larger, and the color center distance is larger (almost the same, except the 4 color centers seeded case) in optimized spherical (ellipsoidal) bare nanoresonators, compared to their coated counterparts.

The average of the short and long axes of ellipsoidal NRs is commensurate with the radius of the spherical NRs, while the shell thickness is ~1.6 (2.2)-times larger, accordingly the $GAR_2$ corresponding to the long axis is commensurate, while the SiV color center's distance is the ~fourth (half) in bare (coated) ellipsoidal NRs, compared to those in their spherical counterpart.

The general characteristic of the optical signals is very similar in NRs with a specific geometry. In all optimized spherical NRs a single peak appears on the extinction and scattering cross-section, on the *Purcell factor* as well as on the *δR* radiative rate enhancement spectrum at the emission wavelength (Fig. 1a-c). In contrast, in the excitation / emission configuration of the optimized bare (coated) ellipsoidal NRs the global maximum appears at (close to) the 532 nm excitation / 737 nm emission wavelength on all spectra, in addition to this a local maximum develops on the extinction cross-section at the excitation wavelength in the emission configuration of coated ellipsoidal NRs.

There is a minimum (local maximum) on the quantum efficiency spectrum at the excitation (emission) wavelength in all spherical NRs (Fig. 1b). The minimum at the excitation is a global minimum in coated spherical NRs. In contrast, a global maximum appears on the *QE* spectrum close to the excitation and at the emission wavelength in corresponding configurations, which is significantly (moderately) larger in the emission configuration of bare (coated) ellipsoidal NRs (Fig. 1b).

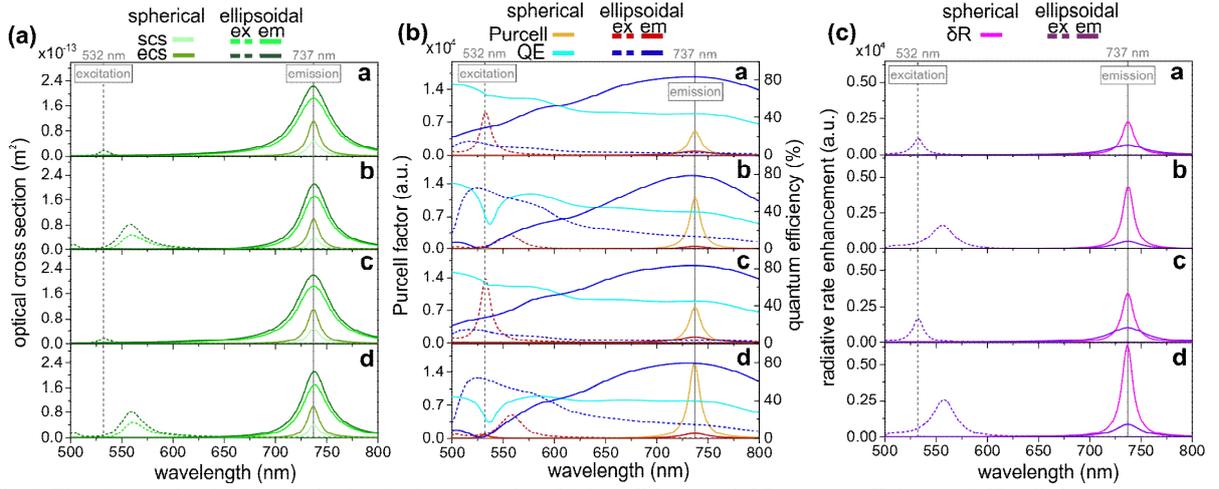

Fig. 1. The (a) scattering (*scs*) and extinction (*ecs*) cross-section, (b) *Purcell factor* and *QE* quantum efficiency, (c) *δR* radiative rate enhancement spectra of (a-c/a) bare_4, (a-c/b) coated_4, (a-c/c) bare_6 and (a-c/d) coated_6 spherical and ellipsoidal NRs.

The time-dependent surface charge density distribution is noticeably (dominantly) hexagonal at the excitation in bare (coated) spherical NRs, while it is permanently dipolar at the emission on all optimized spherical NRs (Fig. 2a and b). In comparison, a noticeable (considerable) hexagonal surface charge density distribution develops at the excitation on bare (coated) ellipsoidal NRs, while a permanently dipolar surface charge density distribution is observable on all optimized ellipsoidal NRs at the emission (Fig. 2c and d). According to the orientation of dipolar emitters in corresponding configurations, the characteristic hexagonal charge distribution is antisymmetric along y (short) axis direction at the excitation, while the dipolar charge distribution is aligned along the z (long) axis direction on the spherical (ellipsoidal) NRs at the emission.

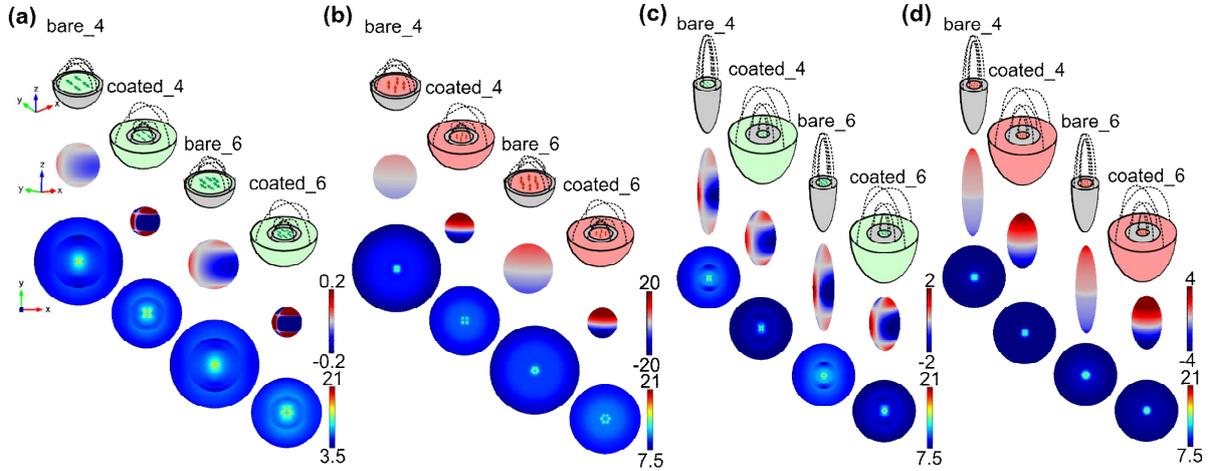

Fig. 2. Schematic drawings, characteristic surface charge density distribution and normalized **E**-field distribution of (a, b) spherical and (c, d) ellipsoidal NRs at the (a, c) excitation and (b, d) emission wavelength.

## 3.1 Comparative study at the excitation

The achieved *QE* is always smaller in NRs consisting of 6 color centers, except in bare ellipsoidal NRs (Fig. 1b, SFig. 2a, STable 2). The larger core, smaller shell thickness and the larger color center distance results in significantly larger *QE* in bare spherical NRs. In contrast, despite the larger long axis, thinner shell and larger (same) emitter distance in case of 4 (6) color centers, the *QE* is significantly smaller in bare ellipsoidal NRs, than in their coated counterparts. Accordingly, the *QE* is significantly smaller (considerably larger) in bare (coated) ellipsoidal NRs, than in their spherical counterparts.

The *Purcell factor* is in the order of 10 and $10^2$ in the optimized bare and coated spherical NRs, while $10^4$ and $10^3$ *Purcell factor* is achieved via bare and coated ellipsoidal NRs (Fig. 1b, SFig. 2b and STable 2). Due to the stronger charge accumulation the *Purcell factor* is always larger in NRs consisting of 6 color centers (SFig. 2b, STable 2). The significantly larger color center distance results in weaker charge accumulation and allows reaching 3 (4)-times smaller *Purcell factor* in bare spherical NRs consisting of 4 (6) color centers, than in their coated counterparts. In contrast, independently of the number of embedded SiV color centers, the commensurate (same) emitter distance makes it possible to achieve stronger charge accumulation and an order of magnitude larger *Purcell factor* in bare ellipsoidal NRs, than in their coated counterparts. Accordingly, two (one) orders of magnitude larger charge accumulation manifests itself in significantly (considerably) larger *Purcell factor* in bare (coated) ellipsoidal NRs, than in their spherical counterparts.

As a result, 4 and 6 (8 and 11)-fold $\delta R_{ex}$ excitation rate enhancement can be achieved in presence of 4 and 6 color centers in bare (coated) spherical NRs. In comparison, $10^3$ ($10^2$)-fold $\delta R_{ex}$ excitation enhancement is achieved in bare (coated) ellipsoidal NRs consisting of either 4 or 6 color centers (Fig. 1c, SFig. 2c, STable 2). Accordingly, the excitation rate enhancement is always larger in NRs consisting of 6 color centers, which manifests itself in larger far-field lobes (Fig. 3a and c). The smaller *Purcell factor* allows smaller excitation rate enhancement and far-field lobes despite the larger quantum efficiency in bare type, than in diamond coated spherical NRs (Fig. 3a). This indicates that the coated spherical NRs can be proposed to achieve non-cooperative SiV color center excitation rate enhancement. In contrast, the one order of magnitude larger *Purcell factor* allows larger $\delta R_{ex}$ excitation rate enhancement and far-field lobes despite the significantly smaller quantum efficiency in bare ellipsoidal NRs compared to their coated counterparts (Fig. 3c). The achieved $\delta R_{ex}$ excitation rate enhancement is significantly (considerably) larger in bare (coated) ellipsoidal NRs, than in their spherical counterparts, which manifest itself in larger lobes corresponding to the far-field radiated power (Fig. 3a-to-c).

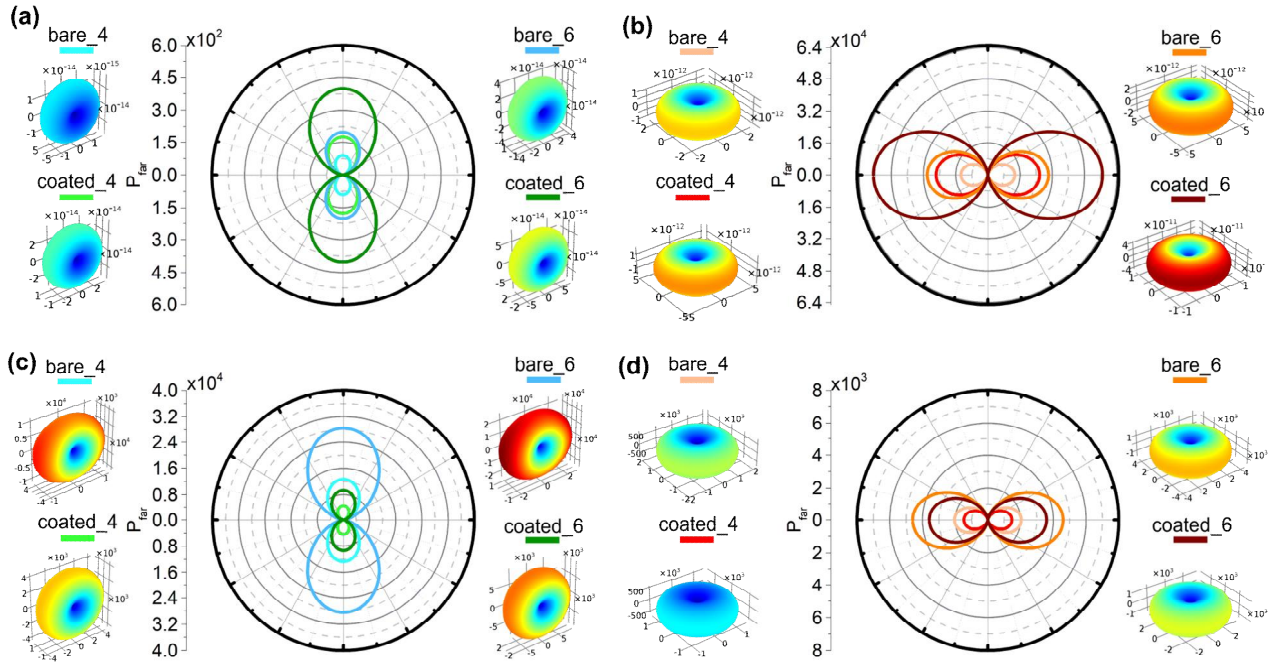

Fig. 3. Angular distribution of the far-field emitted power: (a, b) spherical and (c, d) ellipsoidal NRs at the (a, c) excitation and (b, d) emission wavelength. Insets: schematic drawings and distribution of power emitted into the far-field in 3D. The lobes on the angular distribution of the far-field radiated power are perpendicular to the y (z) axes at the excitation (emission) wavelength, revealing that the SiV color centers are efficiently coupled to each types of optimized concave plasmonic NRs.

## 3.2 Comparative study at the emission

There is no significant difference between the *cQEs* corrected quantum efficiencies achieved in case of seeding by 4 and 6 color centers in neither of spherical nor in ellipsoidal NRs (Fig. 1b, SFig. 2a, STable 2). The larger core (long axis), smaller shell thickness and larger (same) emitter distance makes it possible that the *cQE* at the emission is larger in bare type spherical and ellipsoidal NRs, than in their coated counterparts. The *cQE* is considerably larger both in bare and coated ellipsoidal NRs, than in their spherical counterparts.

Compared to the *QE* at excitation, the *cQE* at the emission is considerably smaller in bare spherical NRs, more commensurate in coated_4 spherical NR, however it is larger in coated_6 spherical NR. In contrast, in all ellipsoidal NRs the *cQE* is larger than the *QE* at the excitation, and the enhancement is significantly larger in bare ellipsoidal NRs.

The *Purcell factor* is in the order of $10^4$ in spherical NRs, while $10^3$ *Purcell factor* is achieved in ellipsoidal NRs, independently of diamond coating existence (Fig. 1b, SFig. 2b, STable 2). The accumulated charge and the *Purcell factor* are always larger in NRs consisting of 6 color centers at the emission as well. The significantly larger (commensurate and same) emitter distance results in weaker charge accumulation in bare spherical (4 and 6 color centers seeded ellipsoidal) NRs, than in their coated counterparts. The smaller amount of charge allows reaching considerably smaller (slightly larger) *Purcell factor* in bare spherical (ellipsoidal) NRs, than in their coated counterparts. The amount of accumulated charge is slightly (one order of magnitude) smaller, which manifests itself in considerably (significantly) smaller *Purcell factor* in bare (coated) ellipsoidal NRs, than in their spherical counterparts.

The accumulated charge (*Purcell factor*) at the emission is typically two (three) orders of magnitude larger than those at the excitation, which reveals that significantly stronger plasmonic resonance occurs at the emission on the coupled spherical NRs. In contrast, the charge accumulation is commensurate at both wavelengths, however it is slightly weaker (stronger) at the emission in bare (coated) ellipsoidal NRs. Accordingly, the *Purcell factor* is smaller with one order of magnitude at the emission in bare ellipsoidal NRs, while it is slightly smaller (larger) in coated_4 (coated_6) ellipsoidal NRs, than that at the excitation. This reveals that the strength of resonance is more commensurate at the two wavelengths on ellipsoidal NRs.

As a result of c*QE* and *Purcell factor*, $2 \times 10^3$ and $4 \times 10^3$ ($4 \times 10^3$ and $7 \times 10^3$) $\delta R_{em}$ emission rate enhancement can be achieved in presence of 4 and 6 color centers in bare (coated) spherical NRs. In comparison, $7 \times 10^2$ and $1 \times 10^3$ ($5 \times 10^2$ and $8 \times 10^2$) $\delta R_{em}$ emission rate enhancement can be achieved in presence of 4 and 6 color centers in bare (coated) ellipsoidal NRs (Fig. 1c, SFig. 2c, STable 2). The radiative rate enhancement and far-field radiated power are always larger in NRs consisting of 6 color centers also at the emission (SFig. 2c, STable 2, Fig. 3b, d). In bare spherical NRs two-times smaller *Purcell factor* allows smaller emission rate enhancement and far-field lobes despite the larger quantum efficiency, similarly to the excitation (Fig. 3b). This indicates again that to achieve non-cooperative SiV color center emission rate enhancement coated spherical NRs can be proposed. In contrast, in bare ellipsoidal NRs the larger *cQE* and larger *Purcell factor* makes it possible to reach larger $\delta R_{em}$ emission rate enhancement and far-field lobes, than that achievable via coated ellipsoidal NRs (Fig. 3d). In contrast to the excitation, the $\delta R_{em}$ emission rate enhancement and the far-field lobes are considerably (significantly) smaller in bare (coated) ellipsoidal NRs, than in their spherical counterparts (Fig. 3b-to-d).

In spherical NRs the radiative rate enhancement at the emission is typically three orders of magnitude larger than that at the excitation, which manifests itself in larger far-field lobes (Fig. 3a-to-b). In contrast, in bare (coated) ellipsoidal NRs the radiative rate enhancement is slightly smaller (larger) than at the excitation, however the maximal extension of the far-field lobe is smaller revealing that the directivity of the coupled system is smaller at the emission in both types of ellipsoidal NRs (Fig. 3c-to-d).

## 3.3 Total fluorescence enhancements

Considerably large $1 \times 10^4$ and $2 \times 10^4$ ($4 \times 10^4$ and $8 \times 10^4$) complete fluorescence enhancement qualified by the $P_x$ *factor* can be achieved with 44% (40%) *cQE* corrected quantum efficiency at the emission in presence of 4 and 6 color centers in bare (coated) spherical NRs. In comparison, $8 \times 10^5$ and $2 \times 10^6$ ($2 \times 10^5$ and $5 \times 10^5$) $P_x$ *factor* can be achieved with 83% (79%) *cQE* corrected quantum efficiency in presence of 4 and 6 color centers in bare (coated) ellipsoidal NRs (Fig. 4a, SFig. 2d, STable 2). Accordingly, $4 \times 10^3$ and $1 \times 10^4$ ($2 \times 10^4$ and $3 \times 10^4$) $P_x cQE$ can be achieved in presence of 4 and 6 color center's in bare (coated) spherical NRs, while $6 \times 10^5$ and $2 \times 10^6$ ($2 \times 10^5$ and $4 \times 10^5$) $P_x cQE$ can be achieved in presence of 4 and 6 color centers in bare (coated) ellipsoidal NRs (Fig. 4a, SFig. 2d, STable 2).

The $P_x$ factor and the $P_x cQE$ are always larger, when 6 color centers are embedded into the NR of the specific type. The larger $P_x$ and $P_x cQE$ achievable in bare_6 and coated_6 NRs, than in their 4 SiV color centers seeded counterparts indicate that both of the total fluorescence enhancement and the FOM converge to larger value, when the number of emitters is increased. However, bare spherical (ellipsoidal) NRs allow reaching smaller (larger) $P_x$ factor and $P_x cQE$ in case of the same number of emitters, according to the smaller (larger) radiative rate enhancement both at the excitation and at the emission. This indicates that bare spherical NRs are less efficient in non-cooperative SiV fluorescence enhancement, while bare ellipsoidal NRs can be proposed for that purpose as well. The fluorescence enhancement qualified by the $P_x$ factor and $P_x cQE$ is two (one) orders of magnitude larger in bare (coated) ellipsoidal NRs, than in their spherical counterparts (Fig. 4a, SFig. 2d, STable 2).

By comparing the achieved total fluorescence enhancements and FOMs, one could already conclude that the ellipsoidal NRs are more efficient to achieve superradiantly enhanced emission. However, from the point of view of superradiance the relative enhancements with respect to the corresponding reference systems provide the most relevant information, which are presented in the next section.

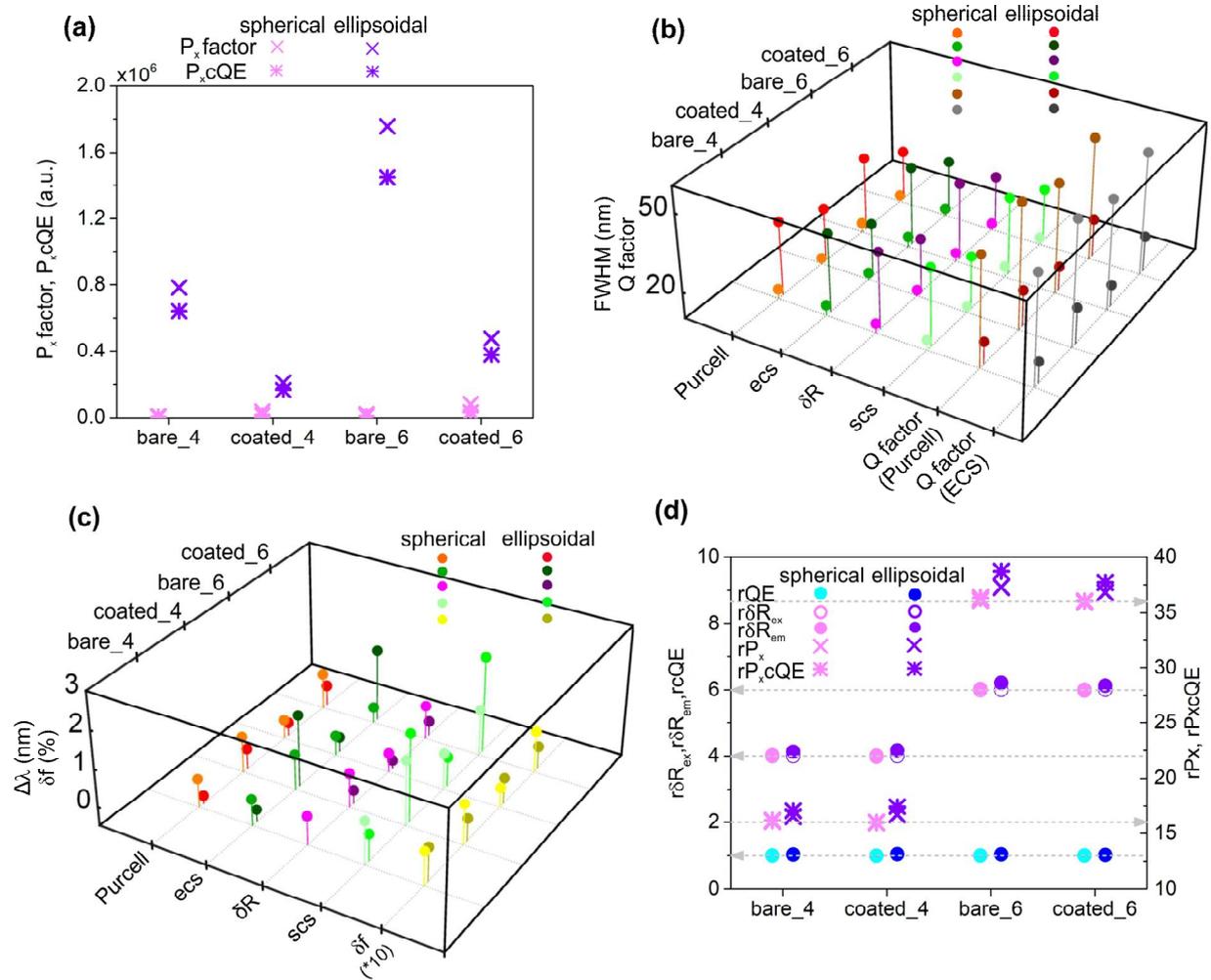

Fig. 4. (a) Comparison of the $P_x$ factor, FOM=$P_x cQE$, $\delta R$ radiative rate enhancement and $cQE$ quantum efficiency to those achieved via corresponding reference systems: proof of superradiance, (b) FWHM of the *Purcell factor* and $\delta R$ radiative rate enhancement, extinction (ecs) and scattering (scs) cross-section peaks and the $Q$ factor computed based on *Purcell factor* and ecs peaks, (c) detuning ($\Delta \lambda$) of the *Purcell factor*, $\delta R$, ecs and scs peaks and relative difference with respect to theoretical frequency pulling ($\delta f$), (d) comparison of the $P_x$ factor and FOM=$P_x cQE$ achieved via optimized spherical and ellipsoidal NRs.

## 3.4 Evaluation of superradiance

Both bare and coated NRs can result in superradiance via 4 and 6 SiV color centers as well, independently of the NR geometry. This can be concluded based on that the balanced partial $N$-fold radiative rate enhancement threshold of superradiance is approximated or reached (is reached or slightly overridden) at the excitation, while it is slightly (considerably) overridden at the emission in spherical (ellipsoidal) NRs (Fig. 4d, SFig. 3a, 3bb and cb, STable 3a-c). The balanced partial $N$-fold excitation rate enhancement threshold of superradiance is just reached in bare_4 spherical, as well as in bare_4 and coated_4 ellipsoidal NRs. The complete multiplied $N^2$-fold $P_x$ factor ($P_x cQE$) threshold of superradiance is overridden in bare_4, bare_6 (as well as in coated_6) spherical NRs, but it is just approximated in coated_4 and coated_6 (coated_4) spherical NRs, while it is overridden by both quantities in all bare and coated ellipsoidal NRs (Figure 4d, SFig. 3bc and cc, STable 3a-c). The $cQE$ at the emission is almost the same in the optimized spherical NRs consisting of either one or multiple color centers, except the bare_6 spherical NR (Figure 4d, SFig. 3ba and ca, STable 3b and c). In contrast, ellipsoidal NRs promote significant $cQE$ improvement as well. All ellipsoidal NRs result in larger relative enhancements, than their spherical counterparts (except the bare_4 at the excitation). Larger number of ellipsoidal NRs overrides the balanced partial $N$-fold $\delta R_{ex}$ threshold of SR at the excitation and more strongly override the balanced partial $N$-fold $\delta R_{em}$ threshold at the emission. Similarly, larger number of ellipsoidal NRs meets the criterion of SR regarding the $P_x$ factor and $P_x cQE$.

The threshold of SR is more significantly overridden in both types of spherical NRs consisting of 6 color centers. The more significant overriding of superradiance threshold (significantly larger values of $P_x$ factor and $P_x cQE$) in both cases of 4 or 6 embedded SiV color centers shows that bare (coated) spherical NRs may have advantages in cooperative (non-cooperative) fluorescence improvement, complementarily. In contrast, the threshold of SR is more (less) significantly overridden in bare (coated) ellipsoidal NRs consisting of 6 color centers (Fig. 4d, STable 3c, SFig. 3cd). These results indicate the advantage of larger number of emitters in both types of spherical and in bare ellipsoidal NRs, but do not confirm the same in case of coated ellipsoidal NRs.

The average of normalized enhancement ratios with respect to the reference system is larger in case of bare spherical NRs independently of the number of emitters. In contrast, it is smaller (larger) in case of bare ellipsoidal NR consisting of 4 (6) color centers (STable 3c, SFig. 3cd). The significantly larger values of $P_x$ factor and the $P_x cQE$ in both cases of 4 and 6 embedded SiV color centers in bare nanoresonators, and the coincident more significant overriding of the superradiance threshold in case of 6 embedded color centers proves the particular advantage of 6 emitters seeded bare ellipsoidal NRs in more efficient superradiantly enhanced fluorescence generation (Fig. 4d, SFig. 2d, STable 2). The better superradiance performance in case of bare composition is unambiguous in spherical NRs and in ellipsoidal NRs consisting of 6 SiV color centers, but it is not confirmed in case of ellipsoidal NRs consisting of 4 color centers.

Ellipsoidal nanoresonators ensure better superradiance performance then their spherical counterparts, independently of the embedded dipole number and NR type.

## 3.5 Evaluation of nanoresonators

The FWHM of the ecs & scs peaks is always larger than that of the *Purcell factor* & $\delta R$ peaks in the optimized NRs, except in coated_4 ellipsoidal NR (Fig. 4b, SFig. 4a, STable 4). This reveals that line-width narrowing occurs in all NRs consisting of cooperatively oscillating emitters, with the one single exception. The FWHM of all inspected spectral peaks only weakly depends on the number of emitters, while the FWHM is uniformly larger in bare NRs than in their coated counterparts both in spherical and ellipsoidal geometries. Moreover, the FWHM of all inspected spectral peaks is larger in ellipsoidal NRs than in their spherical counterparts.

Hereby, one has first to emphasize that to achieve the plasmonic Dicke effect operation in the bad-cavity region is preferred, therefore NRs exhibiting smaller quality factor are proposed. The *Q factor* computed based on the ecs & *Purcell factor* peaks correlate, however the former is slightly smaller in all inspected cases, except in coated_4 ellipsoidal NR. The quality factor is smaller in case of 6 embedded color centers, when it is computed based on ecs and *Purcell factor* peaks of spherical NRs and based on ecs peaks of bare ellipsoidal NRs (Fig. 4b, SFig. 4b, STable 4). The quality factor corresponding to ecs and *Purcell factor* is smaller in bare NRs, than in their coated counterparts, independently of the number of color centers and of the geometry, in accordance with the larger FWHM of corresponding peaks. The quality factor is more than two-times smaller in ellipsoidal NRs than in their spherical counterparts.

Detuning (*Δλ*) of the ecs peaks is commensurate with that of the *Purcell factor* peaks, while detuning of the scs peaks is uniformly larger than that of the *δR* peaks, with one exceptional case of bare_4 ellipsoidal NR (Fig. 4c, SFig. 4c, STable 4). In spherical NRs detuning is smaller in case of 6 emitters, with a few exceptions. In contrast, in bare (coated) ellipsoidal NRs detuning is smaller (larger) in case of 6 emitters, with the one exception of the scs (*δR*) peak. Detuning of peaks in all inspected quantities is smaller in bare spherical NRs than in their coated counterparts. In bare ellipsoidal NRs detuning of the ecs and scs peaks is smaller, while detuning of the *Purcell factor* and *δR* peaks is larger, than in coated ellipsoidal NRs. Ellipsoidal NRs are capable of resulting in smaller detuning of ecs and scs peaks with respect to spherical counterparts, when they are bare, except the ecs peaks in case of 6 embedded color centers. In addition to this, detuning of *Purcell factor* and *δR* peaks is smaller in coated ellipsoidal NRs, than in their spherical counterparts.

The relative difference with respect to the frequency pulling (*δf*) predicted based on theory is smaller (the same) in bare (coated) spherical NRs seeded by 6 color centers (Fig. 4c, SFig. 4d, STable 4). In comparison, *δf* is smaller (larger), when 6 color centers are embedded into bare (coated) ellipsoidal NRs. The difference with respect to the theoretical frequency pulling is smaller (larger) in bare spherical (ellipsoidal) NRs, than in their coated counterparts, independently of the number of color centers. As a result, larger (smaller) relative frequency pulling difference is achievable via bare (coated) ellipsoidal NRs, than via their spherical counterparts.

## 3.6 Applicability in quantum information processing

In spherical NRs the *Purcell factor* indicates a monotonous exponential decay (increase) at the excitation (emission) wavelength, when the distance of the SiV color centers is increased. Degeneracy in the *Purcell factor* at the excitation wavelength is 4 – 4 – 2 and 4 – 2 and 4 –fold, while degeneracy at the emission wavelength is 4 – 4 – 6 – 6 –fold in bare_4 - coated_4 - bare_6 - coated_6 spherical NRs (Fig. 5a, b, SFig. 5).

In contrast, at the excitation in bare_4 (bare_6) ellipsoidal NRs the *Purcell factor* exhibits a non-monotonous distance dependency in all of 4 (a subsystem consisting of 4) SiV color centers, while the sub-system consisting of 2 of the 6 SiV color centers exhibits a monotonous increase in bare_6 ellipsoidal NR. In comparison, at the excitation in coated_4 (coated_6) ellipsoidal NRs the *Purcell factor* decreases exponentially uniformly in all of 4 (differently in subsystems consisting 2 and 4 of the 6) SiV color centers. In emission configuration of ellipsoidal NRs the *Purcell factor* exponentially decreases throughout small emitter distances, however a slow increase is observable in bare ellipsoidal NRs already in the inspected distance interval. Degeneracy in the *Purcell factor* at the excitation is ~4 – 4 – 2 and 4 – 2 and 4 –fold, while degeneracy at the emission is 4 – 4 – 6 – 6 –fold in bare_4 - coated_4 - bare_6 - coated_6 ellipsoidal NRs, similarly to their spherical counterparts (Fig. 5a, b, SFig. 5).

These results indicate that the two subsystems of emitters are distinguishable at the excitation wavelength in 6 color center's seeded spherical and ellipsoidal NRs, while the emitters are indistinguishable in all other NR configurations. The two non-degenerated curves at the excitation converge to the same dependency at distances larger than 20 nm (8 nm) in 6 emitter's seeded spherical (ellipsoidal) NRs. Accordingly, one can conclude that although, the threshold of superradiance is more significantly overridden in case of 6 color centers, the indistinguishability is met at both wavelengths in case of 4 emitters arranged in a square pattern in the NRs caused by symmetry reasons. However, the 2.6% (14.2%) splitting can be neglected in case of bare_6 (coated_6) ellipsoidal, while the larger 56% splitting in 6 color centers seeded spherical NRs has to be taken into account (Fig. 5a, b, SFig. 5).

Both the total fluorescence enhancement and the FOM, namely the $P_x$ and $P_xcQE$ is larger in systems seeded by collectively oscillating emitters than in systems consisting of the same number of randomly oscillating emitters in NRs having the parameters as those determined by optimization realized in presence of emitters with uniform collective phases (Fig. 5c and d). Accordingly, the optimized systems are suitable to achieve superradiantly enhanced fluorescence after synchronization. The difference between $P_x$ *factor* and the $P_xcQE$ achievable via cooperative and randomized systems is significantly more pronounced in ellipsoidal NRs.

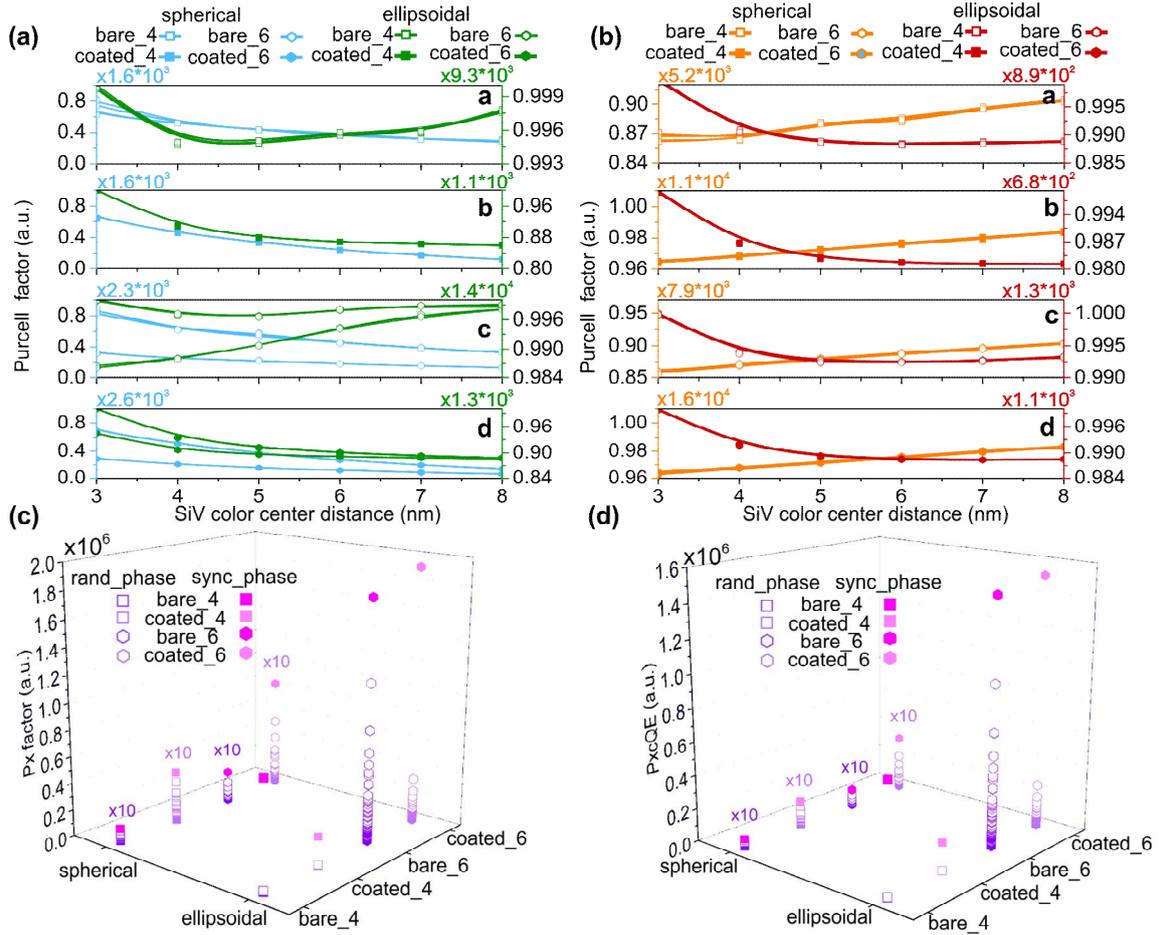

Fig. 5. Degeneracy in *Purcell factor* of the optimized (aa and ba) bare_4, (ab and bb) coated_4, (ac and bc) bare_6, (ad and bd) coated_6 NRs at the (a) excitation and (b) emission wavelengths. Comparison of the $P_x$ *factor* and $P_x cQE$ achieved via (c) spherical and (d) ellipsoidal NRs in case of randomly and collectively oscillating emitters in the optimized NRs.

# 4   Conclusions

Based on our present study in spherical geometry both the $P_x$ *factor* and the $P_x cQE$ are larger in case of 6 color centers, but both are smaller in bare type NRs. In contrast, in ellipsoidal geometry both the $P_x$ *factor* and $P_x cQE$ are larger in presence of 6 color centers and also in bare type NRs. As a result, more seeded bare ellipsoidal NRs have potential to result in more efficient non-cooperative and cooperative fluorescence.

Analysis of cooperative fluorescence has shown that the average of normalized ratios with respect to the reference system is larger in both types of spherical NRs consisting of 6 color centers. In contrast, the sum of normalized ratios with respect to the reference system is larger (smaller) in bare_6 (coated_6) ellipsoidal NRs, than in their 4 color centers embedding counterparts (SFig. 3c, STable 3c).

Although, indistinguishability criterion is met in both configurations of spherical and ellipsoidal NRs consisting of 4 emitters, in NRs embedding 4 color centers the accompanying $P_x$ *factor* and $P_x cQE$ is smaller. Moreover, in bare_4 spherical and ellipsoidal NRs detuning ($\Delta\lambda$) of spectral peaks (except the scs peak in ellipsoidal NRs) and relative difference with respect to theoretical frequency pulling ($\delta f$) is larger, than in their 6 color centers embedding counterparts. In coated_4 spherical (ellipsoidal) NRs detuning of all spectral peaks is larger (smaller) except that of the $\delta R$, while $\delta f$ is equal (smaller), compared to their 6 color centers embedding counterparts. However, via coated_4 spherical NR the bad-cavity criterion is less perfectly met, while in coated_4 ellipsoidal NR there is no line-width narrowing (SFig. 4, STable 4).

The presented results indicate that all characteristics are the less specific to the number of emitters, while the FWHM and the related quality factor correlates with the NR type and geometry, however both $\Delta\lambda$ detuning and $\delta f$ relative difference in frequency pulling exhibit reversal sensitivity to NR type in spherical and ellipsoidal geometry.

One has to emphasize that the relationship between superradiance performance and quality factor may differ in case of different number of emitters, in NRs of different type and geometry, when the maximal FOM is the criterion of coupled system selection. This makes it possible that e.g. coated_4 ellipsoidal NR exhibits better superradiance performance despite the larger quality factor compared to its bare counterpart. Similarly, significantly better superradiance performance is achievable via coated_4 than in coated_6 ellipsoidal NR, even if the quality factor is just slightly smaller.

Taking into account the complexity of the NRs characteristics, the overall ranking of the optimized coupled systems was performed by considering the achieved $P_x$ *factor* and $P_xcQE$, the quality factor, the average of normalized ratios with respect to the reference system qualifying the extent of superradiance threshold overriding as well as the degree of detuning with respect to the SiV color center emission wavelength. This ranking resulted in the coated_4 - bare_4 - coated_6 - bare_6 spherical NR and coated_6 - coated_4 - bare_4 - bare_6 ellipsoidal NR succession. The different superradiance performance and *Q factor* relationship causes the weaker ranking in case of seeding by larger number of emitters of coated ellipsoidal NR in the inspected parameter interval. The resulted ranking indicates that bare type NRs embedding 6 color centers, either of spherical or ellipsoidal geometry, possess the most promising characteristics. Moreover, since all ellipsoidal NRs result in larger average of normalized ratios with respect to the corresponding references systems, than their spherical counterparts, one can conclude that the bare_6 ellipsoidal NR is the most suitable to achieve efficient superradiance.

These results make it possible to select the right seeding, NR type and geometry, which are the most suitable to achieve non-cooperatively or superradiantly enhanced fluorescence, according to the preferences of applications. Based on our present studies, for applications, where complete indistinguishability is important, we propose bare_4 type spherical and ellipsoidal NR exhibiting line-width narrowing with the compromised slightly smaller complete fluorescence enhancement, larger detuning and relative difference in frequency pulling. When partial distinguishability is acceptable, the bare_6 type spherical and ellipsoidal NRs are better, since they exhibit larger fluorescence enhancement, smaller detuning and relative difference in frequency pulling. Moreover, in bare_6 ellipsoidal NR, the threshold of superradiance is more significantly overridden according to the quality factor, which is significantly smaller than the *Q factor* of bare_6 spherical NR.

In summary, concave spherical and ellipsoidal core-shell NRs consisting of larger number of emitters and of bare type are the most suitable to reach superradiance, while coated type spherical and bare ellipsoidal NRs are proposed to achieve non-cooperative fluorescence enhancement. The presented results prove that plasmonic Dicke effect accompanied by complete fluorescence rate enhancement proportional to $N^2$ can be achieved with a balanced radiative rate enhancement proportional to $N$ both at the excitation and emission via optimized plasmonic nanoresonators. Further studies are in progress on different types of more complex plasmonic resonators capable of resulting in superradiantly enhanced emission.

# 5  Acknowledgements

The research was supported by the National Research, Development and Innovation Office-NKFIH through project "Optimized nanoplasmonics" K116362. The project has been supported by the European Union, co-financed by the European Social Fund. EFOP-3.6.2-16-2017-00005. The authors are grateful to Tamás Dániel Juhász for coding of some related programs. Tamás Dániel Juhász was supported by EFOP-3.6.3-VEKOP-16-2017-0002.

## Supporting Information

More details about the applied numerical method, as well as tables and figures about the geometry, optical response and superradiance performance of optimized coupled systems are provided as Supporting Information.